# Nova Centauri 2013 broad maximum from visual observations calibrated with same altitude stars

Costantino Sigismondi

ICRANet Rio de Janeiro, Observatório Nacional Rio de Janeiro, Brazil and AAVSO, American Association of Variable Stars Observers USA.

sigismondi@icra.it

**Abstract:**

A bright Nova in Centaurus was discovered on Dec 2, 2013 at magnitude V=5.5. Its luminosity reached mv=3.8 from Dec 5 to 7, becoming the brightest Nova of 2013 and of the last decades. On Dec 14 it brightened at mv=3.2. The observations of the author contributed to the compilation of the IAU Circular 9265 which announced the Nova as V1369 Cen and are here described. The first observations have been made with naked eye, and the use of same altitude calibration stars of similar color allowed an accuracy within 0.1 magnitudes. The method of selecting stars on the same almucantarat (altitude circle) even farther than 20 degrees, reduces the effects of differential atmospheric extinction. Its application is useful not only for bright Novae but also for luminous variable stars like the supergiant Betelgeuse and Antares and some bright Miras.

**Introduction: Northern vs Southern sky Novae in 2013**

The year 2013 will be recorded as the year of two naked-eye novae: the Nova Delphini 2013 discovered on August 14 which reached mv=4.7 (Sigismondi, 2013; Munari et al., 2013) and the Nova Centauri of December 2nd of mv=3.2 , i.e. 1.5 magnitudes brighter.

The location of the first Nova in the Northern hemisphere and in the summer vacation period allowed an unprecedented observational coverage of such a phenomenon, with almost 500 observers and several thousands of single observations uploaded in the AAVSO database. A large fraction of the observers followed the Nova Del 2013 in the first 10 days after its discovery thanks to the rapid diffusion of the news on the AAVSO website.

Namely 378 observers at Aug. 24, ten days after the discovery, over a total of 490 (at Dec. 17, 2013, data from www.aavso.org ).

The Nova Cen 2013 ten days after the discovery has been observed by 32 people, less than 1/10 of the corresponding number of Nova Del 2013 observers at its first stages. On Dec 17, 15 days after the discovery the total number of Nova Cen 2013 observers is 38. The saturation of the curve describing the evolution of the total number of observers with time seems to be near, because the better observative conditions expected when the Nova will be an object visible for all night will be compensated by its rapid fading, already started.

The first news on this Nova appeared in CBAT (Seach, 2013) on Dec 2nd and it was reported in the AAVSO forum in the same day (Amorim et al., 2013). Up to Dec 7, five days after the discovery, the total number of recorded observers was 24 and the observations uploaded 64. On Dec. 17, the observers are 38, and the observations uploaded 348.

The observers increased after the AAVSO circular #492 (Otero and Waagen, 2013) published on December 4th and the IAU Circular 9265 (Guido et al. 2013) where the name V1369 Cen was assigned to that Nova. The author contributed to that IAUC and to the AAVSO circular with his observations, described hereinafter.

The difference between the number of observers of the two Novae is explainable with the declinations and the time of the night of visibility for the two Novae.

The Nova Del 2013 has a declination of +20º and it was an object culminating at the local midnight at its discovery, easily visible from all continents except Antarctica.

The Nova Cen 2013 has a declination of -59º, visible at its discovery only in the last 2-3 hours of the night. The Nova Cen 2013 is visible only from part of South America and Africa and Oceania. Moreover the Nova Cen 2013 is not high in the sky, from 10º to 40º degrees at its better condition of visibility, occurring only in the final hours of the night: the hardest for amateur observers.

Another aspect to explain the difference in the number of observers between Nova Del and Nova Cen 2013 is demographical, economical and cultural: the number of amateur astronomers is much lower in the Southern hemisphere, even if the brighter magnitude of the Southern Nova would allow a number 3-4 times larger of observers, especially with naked eye thanks to the location of the star very close to β Cen.

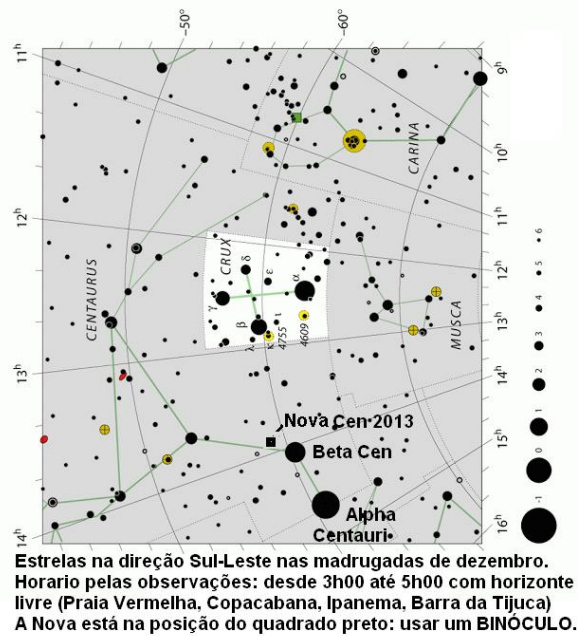

Figure 1: The map to locate the Nova Centauri 2013 prepared for Rio de Janeiro in the occasion of the Day of the Astronomer 7 Dec 2013 (http://www.mast.br/dia_do_astronomo_2013.html ).

On the other hand the presence of the largest observatories in the world in the Southern Hemisphere allows high quality measurements, like spectra and photometry especially when the star will fade (Guido et al., 2013; Izzo et al. 2013).

**Historical aspects:**

The catalogue of the 1022 Fixed Stars of Ptolemy, was considered invariable in number from the medieval commentaries like Sacrobosco, and the stellar variability was a phenomenon recognized only from 1636, with Mira Ceti identification by Holwarda (Sigismondi et al., 2001). Before Mira the occurrence of some Supernovae did not changed the attitude of the western culture towards stellar variability (Thomas, H. L. M., 1948).

Only the Novae were recorded, when particularly bright, in some monastery chronicles (it is the case of SN1006 in Lupus, recorded in San Gallo chronicles).

Chinese observers did record the Novae, or guest stars, with more care than western observers, on behalf of the Emperor and according to their religious beliefs (see Lalande, Histoire de l'Astronomie, 1827). They were more careful in the record of all new stars, guest stars, with or without tail (Ho Peng Yoke, 1962), and the SN 1054 was followed by Chinese astronomers even during the day.

This situation changes with the SN 1572. Tycho Brahe discovered it on November, 11 in the Cassiopeia constellation and it was followed with great observational accuracy, to assess:

1. the existence and the effects of atmospheric refraction

2. the distance of the star, which appeared unaffected by any measurable parallax, so well beyond the planets, located in the eighth sphere, the one of fixed stars.

The SN 1604 of Kepler, was observed on October 19, in Prague and immediately studied by the German astronomer who was there to continue the work of Tycho to 'reshape' astronomy: the observational program of Tycho was published in the Astronomiae Instauratae Mechanica (1598) and in the Astronomia Nova (1609) Kepler used the data of Tycho to interpret the orbit of Mars in heliocentric models.

This Supernova triggered the historical and theological studies of Kepler, who already studied in protestant seminarium, to investigate also the nature of the Star of Bethlehem (e.g. Sigismondi, 2002).

Finally S Andromedae, or SN 1885 according to the following definition of H. Shapley in the 1930s, was the last naked eye Supernova, before the one in the Magellanic Cloud SN 1987A with a visual magnitude around 4 (according to AAVSO Catalogue).

An important role in the study of the Novae was done by the Asiago Observatory from 1942 on (Sigismondi, 2013 and references therein). The Novae have been used as standard candles, like the SN1a are used for cosmological distance estimate of their guest galaxy.

**Anthropological aspects:**

The proper motions of the stars are so small that we see practically the same constellations of several millennia ago. Only precession modifies their coordinate, leaving the asterisms always of the same aspect, but changing their visibility by changing their declinations and their midnight transit time.

So we may not observe stars that our antecessors could see, because they become more austral, like the case of the Alpha and Beta Centauri and the Southern Cross for Southern Europe and North Africa.

The apparition of a Nova may modify the aspect of a constellation for some days or months, as in the case of the Nova Cen 2013, which appeared within one degree from Beta Centauri.

These phenomena may have been remarkable for the ancient cultures, used to observe under very clear skies, and it is confirmed by some archeoastronomical findings (e.g. Navaho drawings in Arizona, as reported by H. Frommert and C. Kronberg, 2006).

**Observational methods: selecting comparison stars on the same almucantarat**

The Nova Centauri 2013 has been observed by the author form the Unisinos (Sao Leopoldo, Porto Alegre - Brazil on December, 4 at naked eye) and from the Centro Brasileiro de Pesquisa Fisica, Rio de Janeiro from December, 5 to 12 with 7x50 binoculars and naked eye. CBPF has been preferred to the Observatorio Nacional location, because of its darker location, protected from the city lights of Rio de Janeiro and having the Atlantic Ocean at South East.

When the Nova Centauri was as bright as 3rd magnitude it was convenient to look for comparison stars at the same altitude of the Nova, i.e. on the same "almucantarat" (arabic name for altitude circle). Reliably their luminosities are affected by the same airmass extinction. Nevertheless there is a rather great spread in the first visual measurements recorded in AAVSO website: are them real or artificial oscillations in the light curve of that Nova?

To orient the naked eye observers the charts of AAVSO do not help because they are limited to a field 1200 arcmin, 20 degrees. They can be used afterwards for photometric references.

With naked eye it is easy to find comparison stars of similar magnitude, but they can be far from the Nova. The program Stellarium (free) helps to recognize the comparison stars, knowing their Hipparcos Catalogue number and its magnitude. With HIP number it is worth to check also the site of SIMBAD http://simbad.u-strasbg.fr/ to verify the V and B magnitudes. A wide field binoculars like 7x50 helps always in doubts and in case of city lights.

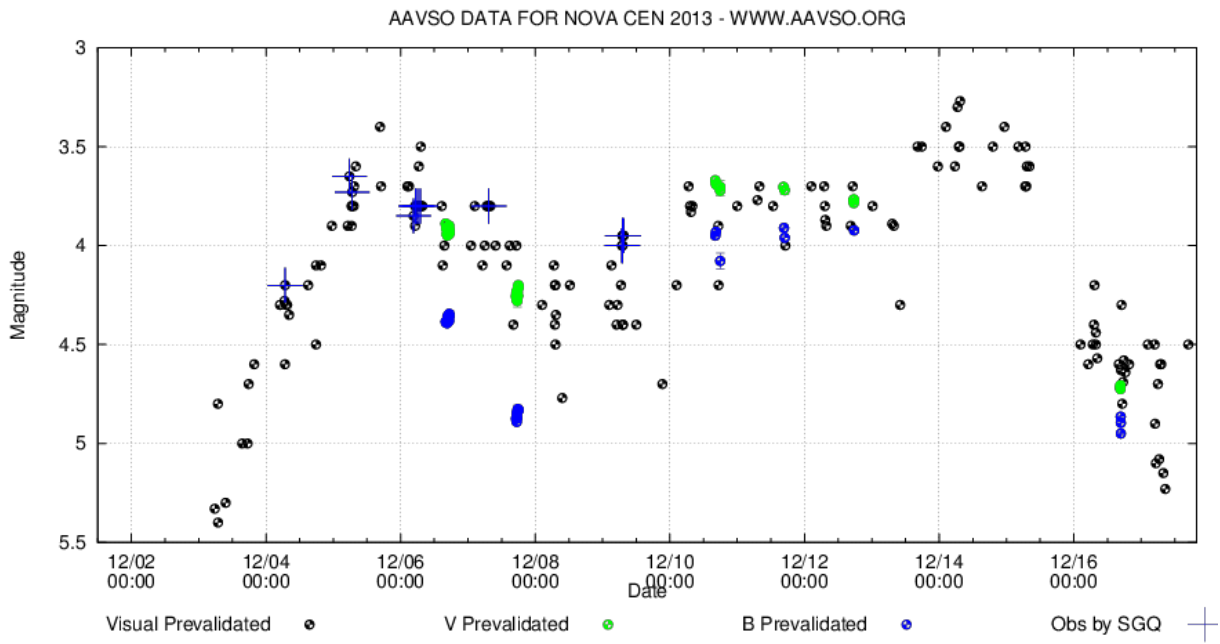

**Figure 1**: The first observations of the Nova Centauri 2013, with the crosses the observations of the author (SGQ AAVSO code). The green and blue spots of Dec. 6-7 are V and B measurements with AAVSO remote telescope (http://www.aavso.org/aavsonet ) made by Arne A. Henden: the B-V =-0.5. Lower values of B-V=-0.2 on Dec 10-11 have been measured by Jonathan Powles in Canberra with another detector; on Dec 16 he showed a rapid luminosity drop suggesting the end of a 10-days bright maximum.

The method of the magnitude estimate is always the method of Argelander, but in order to avoid problems for differential atmospheric extinction, when the comparison stars are far apart, it is recommended to look for same altitude comparison stars. In the case of bright stars they can be as far as 20-30 degrees from the target star.

**Airmass effects for low altitude stars**

I present three examples of airmass extinction:

A) From Rio de Janeiro, Praia Vermelha: because of the Ocean haze and local illumination of the beach I have verified that Gamma TrA mv=3.2 and altitude 15 degrees was at the limiting visibility with naked eye, as well as Epsilon Crucis (mv=3.55) which was at 32 degrees of altitude.

Therefore 0.35 magnitudes were lost on these 17 degrees of altitude from 32 down to 15. If the atmospheric extinction would not be taken into account, there would be an error up to 0.35 magnitudes of the evaluation of the measurement.

B) Even if the standard atmosphere is considered to give an effect of 0.13 magnitudes per airmass (Barbieri, 1999) there are very strong local effects due to relatively thin layers of water vapour, as in the case of sunset measured near the sea (Sigismondi, 2011b). Within 15 arcminutes from the horizon the extinction can be as large as 10-20 magnitudes.

C) The observation of Betelgeuse from Warsaw on Dec 16 at 22:30 local time with full Moon, showed Aldebaran (mv=0.85) Betelgeuse and Rigel (mv=0.15) at the same magnitude. The three stars appeared on three almucantarat respectively of 45, 30 and 15 degrees of altitude. The sky brightened for the full Moon acted with a magnitude loss of 0.7 magnitudes in 30 degrees. The estimate for Betelgeuse was therefore mv=0.5, adding the magnitude loss proportional to its intermediate altitude. The 0.35 magnitudes lost in 15 degrees of altitude are similar to the case A). In this case the Moon was near and at the same altitude of Aldebaran.

"Browsing" the almucantarat: When the Nova is so bright I could see it with naked and unaided eye, and the choice of comparison stars of similar magnitude and altitude was easy thanks to the larger field of view of the eye. I already used this method with delta Scorpii (mv=~1.8, discovered by Sebastian Othero as variable 13 years ago), that from Rome is always very low in the sky, while in Rio de Janeiro it gets also the zenith (Sigismondi, 2011), and with Betelgeuse, that I am following since 2011 for AAVSO.

**Visual estimates within 0.1 magnitudes**

In the night of 9th of December I made several estimates of the Nova using many pairs of comparison stars on the same almucantarat and getting the same result within less than 0.1 magnitudes.

Obviously the higher is the star the better is the estimate, and it happened around 5am of local time: the influence of atmospheric extinction was reduced at minimum.

**Twilight method**

To evaluate the magnitude of the Nova Cen 2013 before sunrise, I followed a method which recalled the heliacal risings recorded by Egyptians and studied by Gaspani and Cernuti (1996). In the morning of Dec 7, 5:30 am, from CBPF, Rio de Janeiro, I saw the pair Beta Cen – Nova Cen 2013 in a hole among the clouds with the Sun only 6º below the horizon, at civil twilight already onset.

The disapparition time of the star (confusion limit with sky background, as observed through the 7x50 binoculars) occurred when the altitude of the star was the same of the previous day: the software of calsky.org was used for calculating the altitudes of the Sun in the two days. The magnitude of the star of Dec 7 was therefore estimated as the same of Dec 6, mv=3.8.

**The distance modulus**

Supposing this Nova at its maximum at mv=3.2, like the Nova Del 2013, and its absolute magnitude of Mv=-9.

We have a distance modulus of 12.2 magnitudes. The correction for galactic reddening and extinction on the line of sight will drive us to real distance modulus. The B-V ranging around -0.2 suggests low values of galactic dust reddening and consequent extinction (Izzo et al., 2013).

The fact that the maximum is much broader than the maximum of the Nova Delphini suggest to wait the luminosity decay rate applying the criterion of Leonida Rosino (see Sigismondi, 2013) to evaluate the absolute magnitude of the maximum of the Nova.

**Concluding remarks and perspectives**

The unaided eye was crucial in the first observations of the Nova Centauri 2013, especially because the majority of the detectors used for variable stars were saturated by such a bright star.

The selection of comparison stars on the same altitude circle, or almucantarat (in arabic historical word) allowed to obtain visual estimates within 0.1 magnitudes of accuracy, exempt from differential atmospheric extinction effects .
When entering such new observations it is recommended to add in the comments "same altitude circle comparison stars", or "same almucantarat comparison stars".

This observational technique is very useful and recommended in case of bright variable stars (Betelgeuse, Antares, the maxima of the brightest Miras) other bright Novae or even for the next Galactic Supernova, that all astronomy community is waiting since 1604 (Bartolini, 2011; Sigismondi, 2005).